\documentclass[prl,aps,twocolumn,showpacs]{revtex4}   
\usepackage{mathrsfs,latexsym,amsmath,psfrag,exscale,pstricks,pst-plot,epsfig}
\usepackage{graphicx}
\usepackage{amssymb}
\newcommand{\be}[1]{\begin{equation}\label{#1}}
\newcommand{\ee}{\end{equation}}     
\newcommand{\bea}{\begin{eqnarray}}
\newcommand{\eea}{\end{eqnarray}}

\begin{document}

\tighten   
%
\title{ \large\bf Small rare gas clusters in soft X-ray pulses}
\author{Christian Siedschlag and Jan M. Rost} 
\affiliation{
 Max-Planck-Institute for the Physics of Complex Systems, 
N\"othnitzer Str. 38, D-01187 Dresden, Germany}
\date{\today}
\begin{abstract}
\noindent
We develop a microscopic model for the interaction of small rare gas
clusters with soft X-ray radiation. It is shown that, while the overall
charging of the clusters is rather low, unexpectedly high atomic charge states
can arise due to charge imbalances inside the cluster. The mechanism does not
require unusually high absorption rates, and the heating can be described by 
standard inverse bremsstrahlung formulae.
\end{abstract}
\draft
\maketitle
\begin{figure}
\centering
\psfrag{a}[][][1]{(a)}
\psfrag{b}[][][1]{(b)}
\psfrag{xtitle}[][][1]{r [a.u.]}
\psfrag{ytitle}[][][1]{E [a.u.]}
\epsfig{file=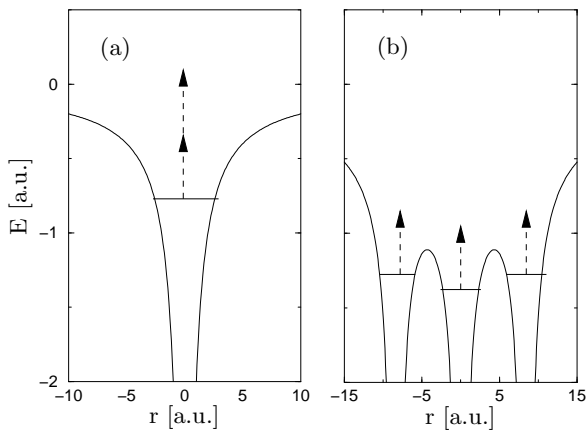, width=0.9 \columnwidth}
\caption{Schematic picture of the (inner) ionization process (a) in a single ion
(e.g. Xe$^+$) and (b) in a simple cluster of three ions. While it takes two photons
to ionize the single ion, one photon is sufficient to achieve inner ionization
in the case of the cluster.}
\label{fig1}
\end{figure}
\begin{figure}
\psfrag{a}[][][1]{(a)}
\psfrag{b}[][][1]{(b)}
\psfrag{c}[][][1]{(c)}
\psfrag{d}[][][1]{(d)}
\psfrag{ytitle1}[][][1]{abs.\ energy [a.u.]}
\psfrag{ytitle2}[][][1]{R(t)}
\psfrag{ytitle3}[][][1]{inner-ionized e$^{-}$}
\psfrag{ytitle4}[][][1]{outer-ionized e$^{-}$}
\psfrag{xtitle}[][][1]{t [a.u.]}
\epsfig{file=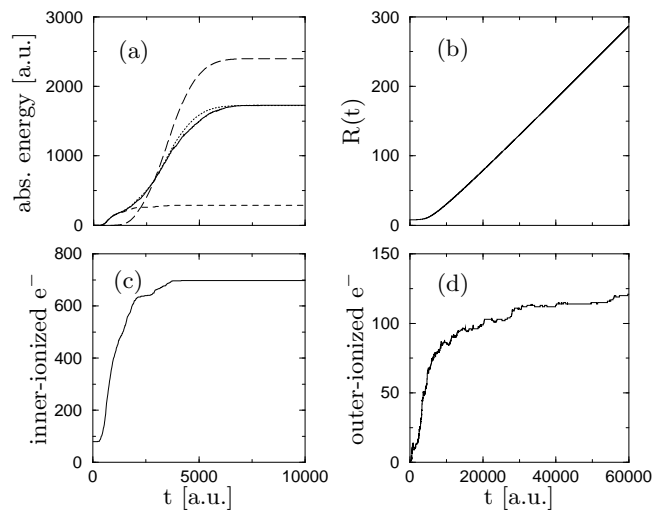,width=0.99\columnwidth}
\caption{Absorbed energy (a), mean interionic distance (b), inner-ionized
electrons (c) and outer-ionized electrons (d) as a function of time for a
Xe$_{80}$-cluster. $I=7\times 10^{13} \ \rm{W/cm^2}$, $T_{\rm{FWHM}}=100 \ \rm{fs}, \omega=12.7$
eV. The different lines in (a) indicate absorbed energy due to inner
ionization, i.e., photo absorption ($E_{ii}$, short-dashed), due to IBS
according to Eq.~\ref{krainov-gl} ($E_{\rm{IBS}}$, long-dashed) and the total energy
absorption from our calculations ($E_{\rm{tot}}$, full line). The dotted line,
 $E'_{\rm{tot}}$, is the absorbed energy constructed from $E_{ii}$ and
$E_{\rm{IBS}}$, see text.}
\label{fig2}
\end{figure}

The interaction of strong lasers with atomic and molecular systems at optical
frequencies has become a vivid field of research, triggering new
developments on both, the experimental and the theoretical side \cite{krainov-review}. For clusters,
an important aspect has been the strongly increased energy absorption
compared to the single atom case. The mechanisms for this phenomenon differ dependent on the size of the cluster. They reach from {\em enhanced ionization} for small clusters \cite{siero}, also known from molecules, over a resonant coupling  of the laser frequency $\omega$ to an collectively
oscillating electron cloud in intermediate clusters \cite{uffe}, to plasma osicllations  whose resonant coupling to the laser frequency are believed to make energy absorption effective in very large clusters \cite{ditmire}.
The order of magnitude of the absorbed energy and  the
observation of highly charged ions and energetic particles could be clearly
reproduced by the corresponding calculations. On the basis of the insight gained so far,  it
seemed clear that in a frequency regime where neither a resonant coupling to the laser frequency is accessible nor the inverse laser frequency is large enough to allow for a
quasistatic description, enhanced energy absorption in clusters should not occur.\\
Hence, it came quite as a surprise when the first XFEL experiment at the DESY in
Hamburg \cite{moeller}, using soft X-ray radiation with $\omega=12.7$ eV and Xenon clusters of up to 30000 atoms, revealed a
complete breakup of the clusters and final ionic charge states of $4+$ for
clusters with $N_{\rm{Atom}} \approx 80$ and up to 8+ when $N_{\rm{Atom}} \approx
30000$, even more, since the maximum intensity was only $I=7 \times 10^{13} \
\rm{W/cm^2}$. At this intensity, single Xenon atoms could only be ionized once
by absorbing a single photon, which is just enough to overcome the
first ionization threshold of Xenon (12.1 eV). Obviously, multiphoton
processes do not play a role.\\
On the other hand, the (surface) plasmon frequency of a spherical nanoplasma with radius
$R$, consisting of $N$
atoms with charge $Z$ and $N_{el}\leq NZ$ electrons, is \cite{deheer}
\begin{align}
\omega_{pl}=\sqrt{NZ/R^3}
\end{align}
which would require charge states of $Z>16$ for the XFEL frequency to come
into resonance with the plasma frequency in the case of a Xe cluster. This
seems extremely unlikely. Moreover, due to the expected expansion of
the cluster, $Z$ would have to be shifted towards even higher values during
the interaction with the pulse. To summarize, none of the concepts from the IR regime
can be taken over to the case of VUV pulses.\\
In this letter we present a theoretical explanation of the observed
phenomenon, based on a mixed quantum-classical model. 
Although the situation is quite different from the IR case, the high charge
states can once again be explained from the fact that a cluster is a dense,
but finite system, i.e., neither an atom nor a quasi-infinite solid or plasma. 
We consider relatively small ($N<100$) Xenon clusters in a short, soft X-ray pulse. The model we have developed is microscopic, allowing to follow all particles and their mutual interactions in time.\\
As it will become clear below, the concept of {\em inner} and {\em outer
ionization},  originally introduced for the case of IR pulses \cite{last} 
is equally useful in the VUV case. {\em Inner ionization} means the
ionization of an electron from  its mother atom (ion), whereas {\em outer
ionization} denotes  the process of an electron leaving the cluster as a
whole. Akin to our previously developed model for rare gas clusters in IR
fields \cite{siero}, we treat the inner ionization process using quantum
mechanical photo absorption rates
while all subsequent time evolution, including outer ionization, is described
by classical mechanics.\\
Whereas in the case of a single Xe atom double ionization with two 
photons
is energetically forbidden, multiple inner ionization can take place in a
cluster due to the influence of the neighbouring ions. The situation is
illustrated in Fig. \ref{fig1}: the interionic barriers are pulled down by the
surrounding charges, so that the difference $\Delta E$ in energy between the level to be
ionized and the top of the nearest barrier becomes less than the energy of one
photon. Consequently, in our treatment an electron can be inner-ionized as long as
$\omega > \Delta E$. If such a photon is absorbed, its energy is transferred
into kinetic energy of the electron (i.e. $|\vec{p}|=\sqrt{2 \omega}$) and the electron
is henceforth treated as a classical particle, starting its trajectory at the
nucleus with momentum $p$. We use specially adapted soft core potentials for
the classical electron-ion-interaction as in \cite{siero} to avoid numerical
singularities and unphysical autoionization processes. The initial 
configuration of the cluster atoms is determined assuming pairwise Lennard-Jones
interaction \cite{LJ}.\\
The probability of an atom or ion absorbing a photon is calculated using
atomic photoabsorption cross sections \cite{rost}. This is justified since these cross
sections can be derived using the reflection principle \cite{refl}, where the electronic
wave function has to be evaluated only once at a distance $r_{\omega}=\omega^{-1/2}$ from
the nucleus. In our case this gives $r_{\omega}=1.46$ a.u., which is small enough to
assume that the atomic wave function at $r_{\omega}$ is hardly changed by the presence of the
neighbouring cluster atoms (at least in the case of van der Waals 
clusters). \\
The inner-ionized electrons, together with the ions created by the ionization
process, form a nanoplasma which is described classically in our
calculation. Here, we do not take into account e-2e processes since the mean
free path for electron-ion-scattering is much larger than the cluster size
\cite{blenski}. Hence, transfer of thermal energy from the nanoplasma electrons to
 electrons still bound is not accounted for. However, the electrons will
transfer thermal energy to the ions causing a slow expansion of the cluster
even if, as a whole, it would stay neutral ({\em hydrodynamic expansion},
cf. \cite{ditmire}).\\
This is of course not the case: The initial energy the electrons get from
the photons is only sufficient for the first few electrons to leave the cluster
before its increasing charge starts to hold them back. The remaining electrons
will evolve into a Maxwellian velocity distribution so that in the course of
time more electrons will gain sufficient kinetic energy to outer
ionize. Moreover, there is the possibility of inverse bremsstrahlung processes
which are included on the classical level: the electrons can absorb energy
from the laser field when undergoing collisions with the ions in the
cluster. We will see later that this process leads to a considerable heating
of the nanoplasma electrons, causing further outer ionization.\\
First of all, however, we shall examine the time dependence of
characteristic observables. Fig.~\ref{fig2} shows the total
energy of the system, the mean internuclear distance, the number of inner-ionized electrons and the number 
of electrons with positive energy (i.e., the number of outer-ionized
electrons). The pulse parameters are as indicated in the figure caption.

\begin{figure}
\psfrag{xtitle}[][][1]{Z}
\psfrag{ytitle}[][][1]{abundance (arb. units)}
\psfrag{a}[][][1]{(a)}
\psfrag{b}[][][1]{(b)}
\epsfig{file=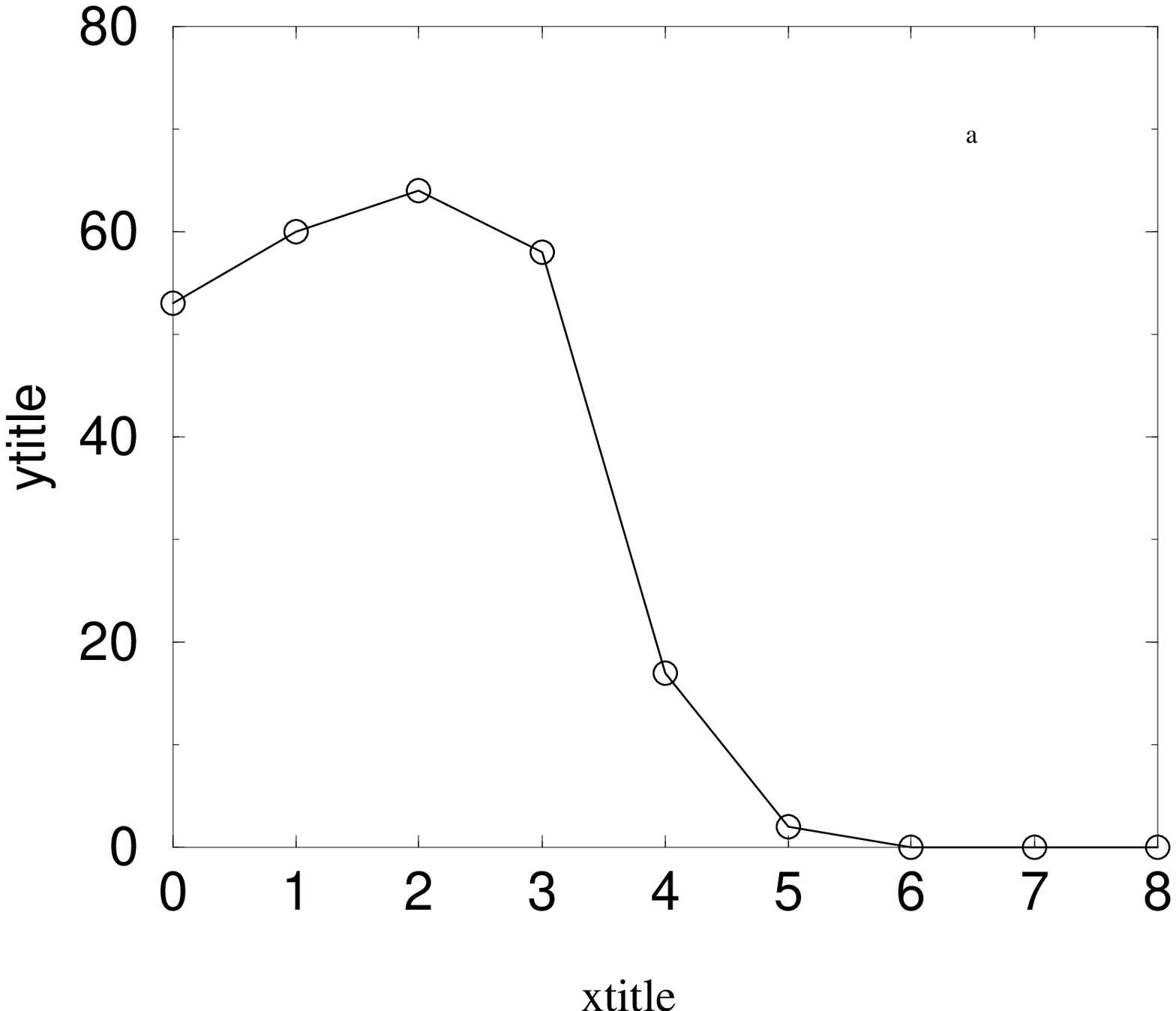,width=0.57\columnwidth,height=0.4\columnwidth}
\epsfig{file=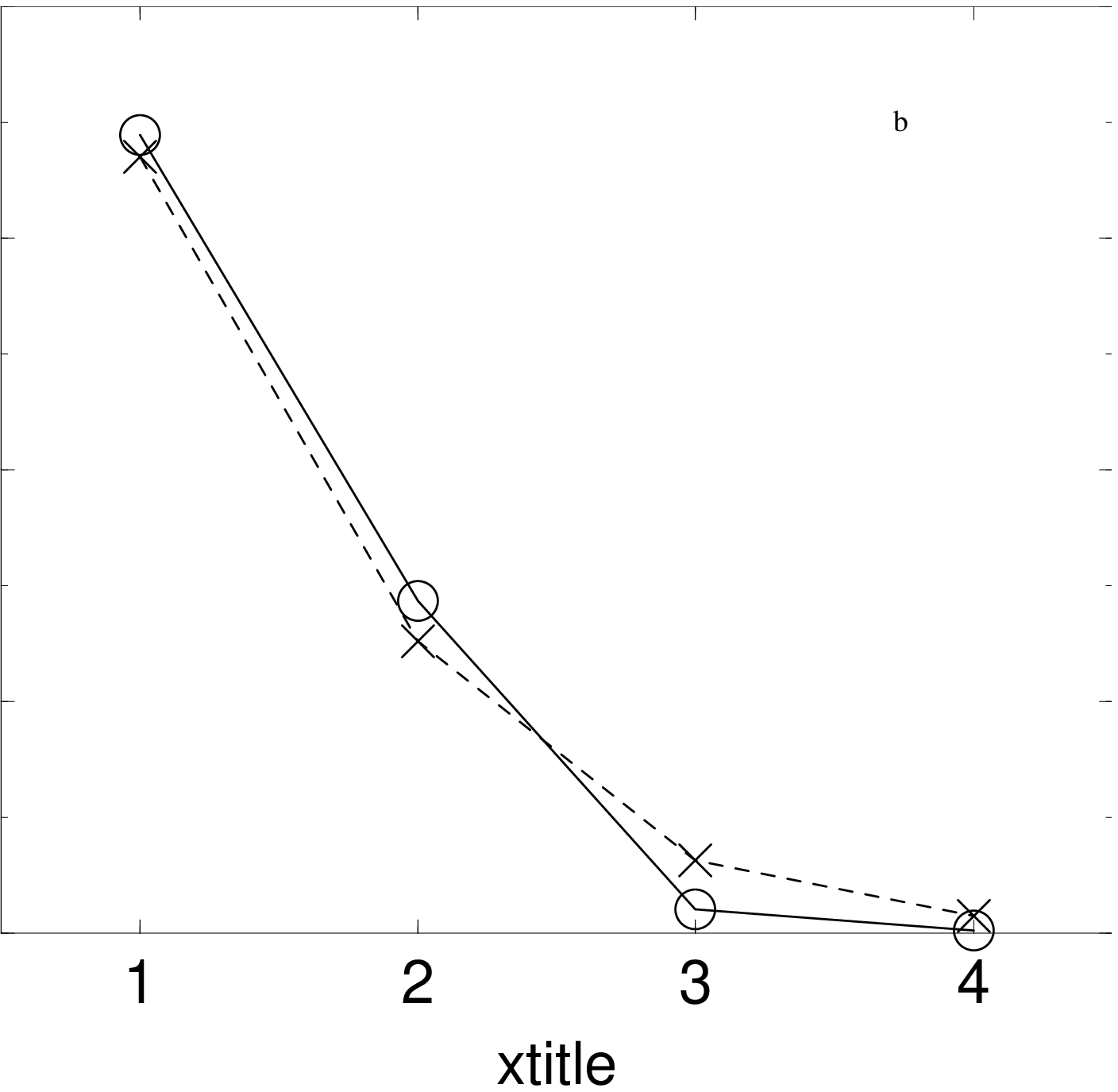,width=0.39\columnwidth,height=0.4\columnwidth}
\caption{Abundance of ionic charge states of Xe$_{80}$ after irradiation from
a soft X-ray pulse: (a) yield in the focus of the pulse, (b) yield integrated
over a gaussian (spatial) pulse profile (circles: our calculations, crosses: measurements from
\cite{moeller}). Pulse parameters are the same as in Fig. \ref{fig2}.}
\label{fig3} 
\end{figure}

We see that the scenario is qualitatively the same as for IR pulses: after the
first electrons have left the cluster, the ions start to repel each other so
that the mean internuclear distance slowly increases, and the cluster
completely disintegrates into ions and atoms. However, the whole process is much less violent in
the VUV case: only about 550 eV per atom are absorbed, at least one order of magnitude less than in IR pulses (however, one order of magnitude more than in the case of an isolated
atom). Due to barrier suppression almost 8 electrons per Xe atom are inner-ionized. The number of outer-ionized electrons
is still slowly increasing well after the end of the pulse, since, due to continuing rethermalization, some electrons can aquire
sufficient kinetic energy to leave the cluster.\\
Note, that the final average charge per atom is only about 1.5. Nevertheless, 
charges up to 5+ can be observed. Before we come to this
point, we must clarify the mechanism of energy absorption. The
cluster can absorb energy in two ways: firstly, if an electron is  inner-ionized the energy of the cluster increases by
$\omega$. Secondly, inner-ionized electrons inside the cluster can absorb energy by
inverse bremsstrahlung (IBS) when colliding with the ions. The
average rate of energy absorption by IBS for a Maxwellian distribution of
electrons with temperature $T$ colliding with ions of charge $Z$ and density $n_i$ in a field of
intensity $I$ and frequency $\omega$ is
\cite{krainov}
\begin{align}
\left<\frac{dE}{dt}\right>=\frac{4 \pi^{3/2}Z^2n_i I}{15 \times 3^{5/6}
\omega^2 \sqrt{2 T}} \left( \frac{2}{Z \omega} \right)^{2/3} \frac{ \Gamma
(1/3)}{\Gamma (2/3)}\,\,\,.
\label{krainov-gl}
\end{align}
In Fig.~\ref{fig2}a we analyze the contributions to energy absorption from 
single photon absorption leading to inner ionization ($E_{ii}$) and from the subsequent IBS
processes ($E_{\rm{IBS}}$). IBS contributes about 80 \% to the absorbed energy
as one can see comparing the full curve $E_{\rm{tot}}$ with the short-dashed
one ($E_{ii}$). $E_{\rm{IBS}}$ can be calculated directly with
Eq. \ref{krainov-gl} taking $T=2/3 \left<E_{\rm{kin}}\right>$ to be the (time-dependent)
temperature of the electrons inside the cluster. The result (long-dashed line)
is too high which can be understood since Eq. \ref{krainov-gl} is strictly valid only when
plasma screening effects can be neglected, which means that the Debye length
$\sqrt{T/(4 \pi Z n_i)}$ should be much larger than the distance between
neigbouring ions. This condition is not fulfilled in our case, so that the
effective ionic charge seen by the electrons is smaller by a certain factor $f$,
or, equivalently, the energy absorption is smaller by a factor $f^2$. With
$f^2=0.6$, one obtains $E'_{\rm{tot}}=f^2\cdot E_{\rm{IBS}}+E_{ii}$ shown as
dotted line in Fig.~\ref{fig2}a in good agreement with the numerical result $E_{\rm{tot}}$. \\
We now turn our attention to the distribution of ionic charges one observes
when the cluster has completely disintegrated. The charge spectrum for
Xe$_{80}$ after being irridiated by the pulse from Fig. \ref{fig2} is shown in
Fig. \ref{fig3}. First of all, charge states of up to 5+ can be observed in
the focus of the pulse (Fig. \ref{fig3} a)). To compare our
results to the measurements in \cite{moeller}, one i) has to integrate over
the spatial intensity distribution of the FEL laser and ii) would have to take
into account the geometrical detector acceptance  for different charge
states, which can only be calculated from the detailed experimental setup and
has not yet been included in the experimental results
\cite{moeller2}. However, with the integration over the Gaussian beam profile
performed, we can at least qualitatively compare our theoretical results with
the experiment, as shown in Fig. \ref{fig3}b. Note that the yield of ions
with charges of five or higher is negligible for Xe$_{80}$ when the volume
integration is performed, which coincides well with the experimental findings.

\begin{figure}
\centering
\psfrag{a}[][][0.8]{a)}
\psfrag{b}[][][0.8]{b)}
\psfrag{c}[][][0.8]{c)}
\psfrag{d}[][][0.8]{d)}
\psfrag{xtitle}[][][1]{r [a.u.]}
\psfrag{ytitle}[][][1]{radial charge densities}
\epsfig{file=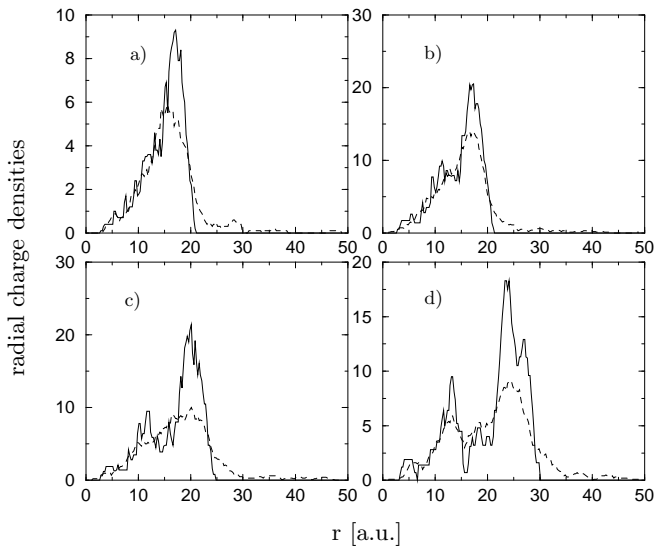,width=0.99\columnwidth}
\caption{Electronic (dashed line) and ionic (solid line) charge densities
$r^2\rho(r)$ at
different times during the interaction: (a) t=500 a.u., (b) t=1500 a.u., (c)
t=3000 a.u, (d) t=4000 a.u. }
\label{fig4}
\end{figure}

The probably most surprising feature are the high charge states, observed in the experiment
and obtained in our calculation.
 In a recent theoretical work \cite{santra} it has been proposed that the high charge states in {\em large} clusters ($N_{\rm{Atom}}>1000$) are due to
thermal equilibration among the particles. While
thermal effects might influence the distribution of the lower charge states,
at least for smaller clusters studied here, the highest charge states arise due to the
finite volume of a cluster which leads to large charge gradients. For simplicity, we will discuss the case of spherical symmetry in the following.
It is known from the laws of electrostatics that a particle with charge $q$ located at a distance $r_q$ from the center of a spherically symmetric charge distribution with 
density $\rho(r)$  interacts with this charge distribution as if it was a point charge at the center of strength $Q(r_{q})=\int_0^{r_q} \rho(r) \ dr$. A 
homogeneous density $\rho(r)=\rho_0$ leads to $Q=\frac 43 \pi \rho_0 r_q^3$. Baring in mind
the $r^{-2}$ dependence of the Coulomb force, an electron inside a positive,
spherically symmetric charge density behaves like a harmonic oscillator (which is just where the surface plasmon
comes from). For this reason, once a few electrons with high kinetic
energy have left the cluster, the remaining electrons will be pulled towards
the center of the cluster, leaving the ions of the outermost shell essentially
stripped \footnote{This requires that the excited cluster indeed behaves like a
spherically symmetric nanoplasma, which we have confirmed by comparing the
radial electronic density with the outcome of a plasma code where
$\rho_{el}(r) \sim \exp{-\phi(r)/kT}$ and $\phi(r)$ is the meanfield potential
calculated from the microscopic eletron and ion density. These results will
be published elsewhere}.\\
This scenario is confirmed by the time evolution of the radial electron and ion
densities  during the pulse. From Fig.~\ref{fig4} it is 
clear that, once the sum of ionic charges exceeds the integrated electron density,
the electrons tend to neutralize the inner part of the cluster (which in this
case consists basically of two ionic shells). Hence, while the outer ionic shells starts already to explode during the time covered by Fig.~\ref{fig4} (charge center of the outer shell at about 20 a.u. in (c) and at more than 25 a.u. in (d)), the inner ionic shell, shielded by the remaining electrons, does not expand, rather its charge center remains at about 12 a.u.. The border between
the neutral and the charged part of the cluster is not completely sharp, as the electrons
possess a finite temperature which broadens their radial distribution, leading
to a smooth distribution of ionic charges.

Hence, our calculations indeed confirm the aforementioned picture for the
emergence of high charge states: it is the finite size of the cluster
which leads to significant charge imbalances inside the cluster, and the
highest ionic charges are therefore generated in the outer shell of the
cluster. For simplicity, we have illustrated this effect with a spherical cluster.
However, any geometric arrangement of the ions which has a clear ''center" will
lead to a strong charge gradient. What is required, however, is a negligible ponderomotive oscillation of
the electrons due to the external field. In the VUV case the oscillation
amplitude is $\sqrt{I}/\omega^2 < 0.2$ a.u. and hence much smaller than the
interionic distances. For IR pulses the electrons are field-driven throughout the
whole cluster, so that any kind of charge imbalance is washed out. On the
other hand, the laser frequency should be small enough to allow for
significant IBS heating: as can be seen from Eq. \ref{krainov-gl}, the rate of
energy absorption decreases quadratically with the applied frequency. This is
why our results should not apply to the case of keV photons \cite{uffe2}.\\
Note that the appeareance of higher charges at the border of the cluster
has to be paid off by some neutral atoms in the center of the cluster (which
also disintegrate due to hydrodynamic expansion). These
neutral atoms, however, could not be detected in the DESY experiment
\cite{moeller2}. In other words, only the relative abundance of {\em ions} has been
measured, with the number of "dark" neutral atoms left unknown. This leads us to
the question of the relevance of our results for the case of larger clusters,
i.e. $N>1000$. We expect the difference between electronic and ionic density
at the surface of the cluster to be even bigger than for small clusters, since
the space charge will be higher too. Thus, the maximum charge state attainable
by charge imbalance will increase with the cluster size. On the other hand,
the relative proportion of the surface region to the total volume of the
cluster will decrease, leaving relatively more neutralized ions in the center
of the cluster. As mentioned, these neutral atoms would have been invisible in
the experiment; the {\em relative} abundance, however, might well be
reproducable by our model even for larger clusters. Moreover, a new set of
experiments is planned at DESY where, e.g., Argon clusters can be doped with a
single Xenon atom which can be placed in the center of the cluster or at the
surface, respectively \cite{moeller2}. This kind of experiment could directly
verify the predictions made by our model.\\
To summarize, we have presented a microscopic model which simulates the
time evolution of rare gas clusters in VUV laser fields. Single-photon processes, leading
only to single ionization for isolated Xenon atoms, can create a dense nanoplasma in
the case of a cluster because inner ionization is facilitated by the charged
environment of a
cluster atom, effectively reducing the threshold for ionization. Once the
nanoplasma is formed, the energy absorption and
final average ionic charge turn out to be within the limits set by standard
inverse bremsstrahlung theory; nevertheless, the geometric properties of
clusters lead to unusually high charge states originating from the surface
region of the cluster. \\
We thank Ulf Saalmann and Thomas Pohl for helpful discussions.

\bibliographystyle{unsort}

\end{document}